\documentclass[pre, a4paper, superscriptaddress, amsfonts, amssymb, amsmath, reprint, showkeys, nofootinbib, oneside]{revtex4-1}
\usepackage[english]{babel}
\usepackage[utf8]{inputenc}
\usepackage{xcolor}
\usepackage{bm}% bold math
\usepackage{graphicx}% Include figure files
\usepackage{float}
\usepackage{amsmath}
\usepackage{amsfonts,amssymb,amsthm}
\usepackage[colorinlistoftodos, color=green!40, prependcaption]{todonotes}
\usepackage{amsthm}
\usepackage{mathtools}
\usepackage{physics}
\usepackage{xcolor}
\usepackage{graphicx}
\usepackage[left=23mm,right=13mm,top=35mm,columnsep=15pt]{geometry} 
\usepackage{adjustbox}
\usepackage{placeins}
\usepackage[T1]{fontenc}
\usepackage{lipsum}
\usepackage{csquotes}
\usepackage[pdftex, pdftitle={Article}, pdfauthor={Author}]{hyperref} % For hyperlinks in the PDF
\usepackage{notes2bib}

\begin{document}
\title{Dissecting Spectral Granger Causality through Partial Information Decomposition}

\author{Luca Faes}
    \email[Correspondence email address: ]{luca.faes@unipa.it}% Your name
    \affiliation{Department of Engineering, University of Palermo, Italy}
    \affiliation{Faculty of Technical Sciences, University of Novi Sad, Serbia} 
\author{Gorana Mijatovic}
    \affiliation{Faculty of Technical Sciences, University of Novi Sad, Serbia}
\author{Riccardo Pernice}
    \affiliation{Department of Engineering, University of Palermo, Italy}
\author{Daniele Marinazzo}
    \affiliation{Department of Data Analysis, University of Ghent, Belgium} 
\author{Sebastiano Stramaglia}
    \affiliation{Department of Physics, University of Bari Aldo Moro, Italy}    
\author{Yuri Antonacci}
    \affiliation{Department of Engineering, University of Palermo, Italy}

\begin{abstract}
Granger causality (GC), a popular statistical method for the inference of directional influences between time series measured from a complex network, is sensitive to high-order (non-pairwise) interactions which fundamentally shape the collective network dynamics. This work introduces Partial Decomposition of Granger Causality (PDGC), a tool eliciting redundant and synergistic causal interactions in the pattern of information flow between the subsystems of physiological networks. The tool exploits the framework of partial information decomposition to dissect the multivariate GC from a set of driver random processes to a target process into unique effects carried exclusively by each driver, redundant effects carried identically by more drivers, and synergistic effects carried jointly by some drivers but not by any of them individually. Computation is based on multivariate state-space models expanded in the frequency domain to assess PDGC both in specific bands of physiological interest and in the time domain after whole-band integration. 
The validation on benchmark simulations demonstrates that the measures of unique, redundant, and synergistic GC reflect the underlying causal mechanisms and are computationally reliable. The application to arterial pressure, respiration, cerebral blood velocity and heart period variability reveals striking differences in the response to postural stress of patients prone to neurally-mediated syncope compared to healthy controls. The extraction of high-order causality patterns from the spectral GC favors dissecting the mechanisms of causal influence underlying multivariate interactions among oscillatory processes in many data-driven applications of network science. 
\end{abstract}

%\keywords{first keyword, second keyword, third keyword}

\maketitle

\section{Introduction} \label{sec:introduction}
Understanding the functional structure and interactions of complex dynamical systems formed by multiple interacting units is a main challenge in in the field of network physiology \cite{lehnertz2020human}. In this context, Granger Causality (GC) is a well-grounded, very popular tool for inferring directional interactions between pairs of network units by accounting for the flow of time \cite{porta2015wiener}. GC is rooted in the notion of time-lagged influence formalized by Granger \cite{granger1969investigating} and Geweke \cite{geweke1982measurement} in the context of autoregressive (AR) modeling of random processes. The popularity of GC stems from its data-driven nature, its solid information-theoretic foundations \cite{barnett2009granger}, and the availability of multivariate and conditional forms, together with their corresponding spectral representations, which allow to identify causal influences focusing on direct connections and on specific frequency bands of interest \cite{geweke1982measurement,baccala2001partial}. Thanks to these properties, GC is widely used in several fields, including the analysis of physiological oscillations in a variety of patho-physiological conditions \cite{porta2013cardiovascular,stramaglia2016synergetic,porta2023concomitant,sparacino2023method,pernice2022spectral}.

Another rapidly growing line of research in network science investigates how collective dynamics emerge out of the causal links wiring the nodes of a network system \cite{rosas2020reconciling,rosas2022disentangling,faes2025predictive}. These emergent behaviors, which are typically denoted as high-order interactions (HOIs), represent interaction modes that involve the dynamic activity of more than two nodes and thus cannot be traced by simple pairwise analyses. Accordingly, several HOI measures quantifying the redundant or synergistic nature of the information shared in groups of network units mapped by random variables have been developed along this line \cite{williams2010nonnegative,rosas2019quantifying,varley2023multivariate}. The recent generalization of these measures to random processes with temporal statistical structure \cite{faes2022new, faes2025partial} allowed to account for dynamical information flows in the detection of HOIs and favored the application to networks of brain and physiological oscillations \cite{scagliarini2024gradients,faes2025predictive}.

Even though they are typically used to investigate different aspects of the functional structure of network systems (respectively, directed information flow and redundant/synergistic statistical dependencies), the notions of GC and HOIs are interrelated and the measures used to quantify them share common information-theoretic grounds. In fact, the existing multivariate formulations of GC \cite{geweke1984measures,barrett2010multivariate,porta2015wiener} investigate multiple time series and can thus be inherently regarded as high-order tools. However, although previous works have explored the effects of HOIs on GC showing that pairwise and conditional GC are sensitive respectively to redundancy and synergy \cite{stramaglia2014synergy,stramaglia2016synergetic}, an explicit link between measures of GC and quantities directly reflecting HOIs has not been established yet.

To fill this gap, the present study introduces the so-called partial decomposition of Granger causality (PDGC), a tool dissecting the multivariate GC from several driver processes to a target process into components quantifying the unique causal effects exerted over the target by each individual driver, as well as separate redundant and synergistic high-order causal effects. 
PDGC embeds time- and frequency-domain GC measures computed from the state-space (SS) modeling of multiple time series \cite{barnett2015granger} into the framework of partial information decomposition (PID) \cite{williams2010nonnegative}. Exploiting a frequency-specific definition of redundancy \cite{faes2025partial}, the PID of the multivariate GC is defined both in the spectral domain to dissect HOIs related to specific oscillations embedded in the analyzed processes, and in the time domain after spectral integration.

\textcolor{black}{After illustrating their theoretical behavior and computational reliability in benchmark simulations}, the PDGC measures are applied to a network widely studied to characterize the short-term homeostatic regulation of the human organism, i.e. the network of cerebrovascular, cardiovascular and respiratory interactions \cite{gelpi2022dynamic,pernice2022spectral}. Specifically, we aim to elucidate the neurophysiological mechanisms underlying the  beat-to-beat control of heart rate and cerebral blood flow, investigating how their oscillations are influenced by arterial pressure and respiration variability in healthy subjects and in patients prone to the development of postural-related syncope.
\textcolor{black}{The PDGC measures provided by the framework are collected as part of the HOP Matlab toolbox, described in the supplemental material of this article and freely available for download at www.lucafaes.net/hop.html  and https://github.com/YuriAntonacci/HOP.}

\section{Methods} \label{sec:Methods}

\subsection{Granger causality} \label{sec:GC}
%Granger causality (GC) is typically formulated in terms of prediction error resulting from regression models \cite{granger1969investigating}. In this section, following the formulation of Geweke \cite{geweke1982measurement}, we provide formal definitions of GC in both time and frequency domains.

\subsubsection{Time-domain formulation}
Let us consider a discrete-time, real-valued $M$-dimensional random process $S=\{S_n\}_{n\in \mathbb{Z}}$, $S_n \in \mathbb{R}^{M}$, which is partitioned into a scalar target process and $N=M-1$ driver processes: $S_n=\{X_n,Y_n\}$, $X_n\in\mathbb{R}^{N}$, $Y_n \in \mathbb{R}$.
In this frame, the best least-squares prediction of the present state of the target process, $Y_n$, given the entire past information retrieved from the whole process, $S_n^-=[S_{n-1}^\intercal S_{n-2}^\intercal\cdots]^\intercal$, is the projection of $Y_n$ on the space spanned by $S_n^-$, i.e. $\mathbb{E}[Y_n|S_n^-]$, and the prediction error $U_{Y|S,n}=Y_n-\mathbb{E}[Y_n|S_n^-]$ is a white noise process with variance $\sigma^2_{Y|S}=\mathbb{E}[U_{Y|S,n}^2]$. This prediction is contrasted with the prediction of the present state of the target only based on its own past information, $\mathbb{E}[Y_n|Y_n^-]$, yielding a prediction error $U_{Y|Y,n}=Y_n-\mathbb{E}[Y_n|Y_n^-]$ with variance $\sigma^2_{Y|Y}=\mathbb{E}[U_{Y|Y,n}^2]$.
%$Y_n$ based on the past information retrieved from the \textit{reduced} process involving only the driver $X$, $\mathbb{E}[Y_n|X_n^-]$, which yields a prediction error $U_{Y|X,n}=Y_n-\mathbb{E}[Y_n|X_n^-]$ with variance $\sigma^2_{Y|X}=\mathbb{E}[U_{Y|X,n}^2]$.
Then, the time-domain GC from $X$ to $Y$ is defined as \cite{geweke1982measurement}:
\begin{equation}
    F_{X\rightarrow Y}=\ln \frac{\sigma^2_{Y|Y}}{\sigma^2_{Y|S}},
    \label{GCtime}
\end{equation}
quantifying the predictive information brought by the past states of the driver process to the present state of the target above and beyond its self-predictability.

Crucially, GC can be computed similarly by taking a subset of $N_j<N$ driver processes identified by the indices $\alpha_j=\{j_1\cdots j_{N_j}\}\subset \{1\cdots N\}$; the corresponding sub-process is $X_{\alpha_j}=\{X_{j_1},\ldots, X_{j_{N_j}}\}$.
%$X_{\alpha_j} \subset X$; when referring to a sub-process, we use the notation $X_{\alpha,n}:=[X_{j_1,n}\cdots X_{j_K,n}]\in \mathbb{R}^K$, with $\{j_1,\ldots,j_K\}\subset \{1,\ldots,N\}$, $K<N$.
Considering the restricted process $Z=\{X_{\alpha_j},Y\}\subset S$ formed by the variables $Z_n=[X_{\alpha_j,n}^\intercal Y_n]^\intercal \in \mathbb{R}^K$, $K=N_j+1$, from which the prediction of $Y_n$ based on the past information contained in $Z$ yields a prediction error $U_{Y|Z,n}=Y_n-\mathbb{E}[Y_n|Z_n^-]$ with variance $\sigma^2_{Y|Z}=\mathbb{E}[U_{Y|Z,n}^2]$, the GC from $X_{\alpha_j}$ to $Y$ is given by:
\begin{equation}
    F_{X_{\alpha_j}\rightarrow Y}=\ln \frac{\sigma^2_{Y|Y}}{\sigma^2_{Y|Z}}.
    \label{GCtime_sub}
\end{equation}
Note that, while the pairwise GC (\ref{GCtime_sub}) quantifies the causal influence of $X_{\alpha_j}$ on $Y$ without explicitly considering the effects of the remaining processes collected in $X^{\alpha_j} = X \backslash X_{\alpha_j}$, a so-called conditional form of the GC can be obtained simply as $F_{X_{\alpha_j}\rightarrow Y|X_{\alpha_j}^-}=F_{X\rightarrow Y}-F_{X^{\alpha_j}\rightarrow Y}$ \cite{barrett2010multivariate}.

The most popular implementation of GC is based on the vector autoregressive (VAR) representation of the process $S$:
\begin{equation}
   \textcolor{black}{S_n} = \sum_{k=1}^{p}\mathbf{A}_k S_{n-k} + U_n
   \label{VARmodel},
\end{equation}
where $p$ is the VAR model order, the matrix $\mathbf{A}_k \in \mathbb{R}^{M \times M}$ contains the coefficients relating the present with the past of the processes at lag $k$, and $U_n$ is a white noise process with positive-definite covariance matrix $\boldsymbol{\Sigma}_U=\mathbb{E}[U_nU_n^\intercal]$; such matrix contains the variance $\sigma^2_{Y|S}$ as the $M^{\mathrm{th}}$ diagonal element.
Besides the full VAR model (\ref{VARmodel}), to compute GC it is necessary to identify reduced models describing subsets of processes. The VAR model representation of the reduced process $Z=\{X_{\alpha_j},Y\}$ is given by:
\begin{equation}
   \textcolor{black}{Z_n} = \sum_{k=1}^{\infty}\mathbf{B}_k Z_{n-k} + W_n
   \label{redVARmodel},
\end{equation}
where $\mathbf{B}_k \in \mathbb{R}^{K \times K}$ and $W_n \in \mathbb{R}^K$ are the coefficients and the noise innovations. The innovation covariance matrix, $\boldsymbol{\Sigma}_W=\mathbb{E}[W_nW_n^\intercal]$, contains the variance $\sigma^2_{Y|Z}$ as the $K^{\mathrm{th}}$ diagonal element. If $\alpha_j=\{\emptyset\}$, (\ref{redVARmodel}) reduces to an AR model of the scalar target process $Y$, with innovation variance $\sigma^2_W=\sigma^2_{Y|Y}$.

\subsubsection{Frequency-domain formulation}
To compute GC in the frequency domain, we adapt the approach proposed in \cite{geweke1982measurement} to our setting, first partitioning the VAR sub-model (\ref{redVARmodel}) as:
\begin{subequations}
\begin{align}
    X_{\alpha_j,n} &= \sum_{k=1}^{\infty}\mathbf{B}_{11,k} X_{\alpha_j,n-k} + \sum_{k=1}^{\infty}\mathbf{B}_{12,k}Y_{n-k} + W_{1,n} \label{redVAR2}\\
    Y_n &= \sum_{k=1}^{\infty}\mathbf{B}_{21,k}X_{\alpha_j,n-k}+ \sum_{k=1}^{\infty}b_{12,k}Y_{n-k} + W_{2,n}, \label{redVAR1}
\end{align} \label{redVARdec}
\end{subequations}
to evidence the VAR representations of the driver and target processes, which depend on the noise innovations $W_{1,n} \in \mathbb{R}^{N_j}$ and $W_{2,n}\in \mathbb{R}$ with covariances $\Sigma_{1}=\mathbb{E}[W_{1,n}W_{1,n}^\intercal]$ and $\sigma^2_2=\mathbb{E}[W^2_{2,n}]$. The two innovations are generally correlated ($\Sigma_{12}=\Sigma_{21}^\intercal=\mathbb{E}[W_{1,n} W_{2,n}]\neq 0$) due to the instantaneous interactions between the drivers and the target. To remove these zero-lag effects from the spectral representation of the time-lagged GC from $X_{\alpha_j}$ to $Y$, the Cholesky decomposition is applied \cite{geweke1982measurement,faes2013framework} pre-multiplying (\ref{redVARdec}) by $[\mathbf{I}_{N_j} -\frac{\Sigma_{12}}{\sigma^2_2}]$ to obtain:
\begin{equation}
    X_{\alpha_j,n} = \sum_{k=1}^{\infty}\tilde{\mathbf{B}}_{11,k} X_{\alpha_j,n-k} + \sum_{k=0}^{\infty}\tilde{\mathbf{B}}_{12,k}Y_{n-k} + \tilde{W}_{1,n},
    \label{redVAR2tilda}
\end{equation}
which includes instantaneous effects between $Y_n$ and $X_{\alpha_j,n}$ quantified by the coefficients $\tilde{\mathbf{B}}_{12,0}$, and has innovations $\tilde{W}_{1,n} \in \mathbb{R}^{N_j}$ with covariance $\tilde{\Sigma}_{1}=\Sigma_{1}-\frac{\Sigma_{12}\Sigma_{21}}{\sigma^2_2}$.
Then, the VAR model with instantaneous effects formed by (\ref{redVAR2tilda}) and (\ref{redVAR1}), which has uncorrelated innovations ($\tilde{\Sigma}_{12}=\mathbb{E}[\tilde{W}_{1,n} W_{2,n}]=0$), is studied in the frequency domain by taking its Fourier transform (FT) to get $Z(\omega)=\tilde{\mathbf{H}}(\omega)\tilde{W}(\omega)$, where $Z(\omega)=[X_{\alpha_j}(\omega)^\intercal Y(\omega)]^\intercal$ and $\tilde{W}(\omega)=[\tilde{W}_1(\omega)^\intercal W_2(\omega)]^\intercal$ are the FTs of $Z_n$ and $\tilde{W}_{n}=[\tilde{W}_{1,n}^\intercal W_{2,n}]^\intercal$, and $\tilde{\mathbf{H}}(\omega)$ is the $K \times K$ transfer function (TF) matrix
($\omega=2\pi\frac{\bar{f}}{\bar{f}_s} \in [0,\pi]$ is the normalized angular frequency, with $\bar{f}$ frequency and $\bar{f}_s$ sampling frequency). The TF matrix is used to compute the $K \times K$ power spectral density (PSD) matrix of the process $Z$ as:
\begin{equation}
    \mathbf{P}_Z(\omega)=\tilde{\mathbf{H}}(\omega)\Sigma_{\tilde{W}}\tilde{\mathbf{H}}^*(\omega),
    \label{PSDfact}
\end{equation}
where $\Sigma_{\tilde{W}}=\mathbb{E}[\tilde{W}_n \tilde{W}_n^\intercal]$. Both $\mathbf{P}_Z(\omega)$ and $\tilde{\mathbf{H}}(\omega)$ can be factorized in diagonal blocks of dimension $N_j \times N_j$  and $1 \times 1$ evidencing the PSDs and internal TFs of the processes $X_{\alpha_j}$ and $Y$, and off-diagonal blocks of dimension $N_j \times 1$ and $1 \times N_j$ evidencing the cross-PSDs and cross-TFs between  $X_{\alpha_j}$ and $Y$. Using this factorization in (\ref{PSDfact}) and the fact that $\Sigma_{\tilde{W}}$ is block-diagonal, the PSD of the target process can be written as 
\begin{equation}
    P_Y(\omega)=\tilde{H}_{21}(\omega)\tilde{\Sigma}_{1}\tilde{H}_{21}^*(\omega)+\sigma_2^2|\tilde{H}_{22}(\omega)|^2,
    \label{PSDY}
\end{equation}
from which the spectral GC from $X_{\alpha_j}$ to $Y$ is defined as \cite{geweke1982measurement}:
\begin{equation}
    f_{X_{\alpha_j} \rightarrow Y}(\omega)=\ln\frac{P_Y(\omega)}{\sigma_2^2|\tilde{H}_{22}(\omega)|^2}.
    \label{SGC}
\end{equation}
The frequency-domain formulation of the GC provided in (\ref{SGC}) has several desirable properties, including its non-negativity and the fact that it can be closely related to the time-domain GC by the spectral integration property, i.e. $F_{X_{\alpha_j}\rightarrow Y}=\frac{1}{\pi}\int_0^{\pi}f_{X_{\alpha_j} \rightarrow Y}(\omega)\mathrm{d}\omega$. 
Remarkably, the definition (\ref{SGC}) differs from the spectral causality measure computed in \cite{faes2022new} in the way instantaneous effects among the processes in $X_{\alpha_j}$ are handled; the two measures are equivalent when the sub-process $Z$ is strictly causal ($\Sigma_{12}=\Sigma_{21}=0$); see the Supplementary Material (Sect. II.C) for a detailed comparison.

\subsection{Partial Decomposition of Granger Causality} \label{sec:HOGC}
\begin{figure} [t!]
    \centering
    \includegraphics[scale= 1.15]{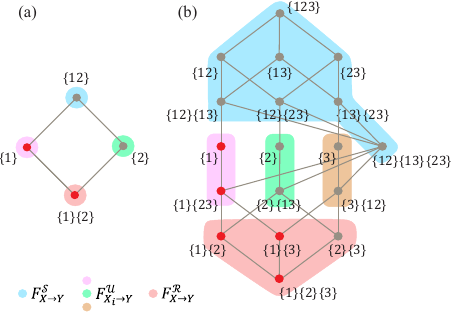}
    \caption{Granger Causality redundancy lattices for $N$=2 (a) and $N$=3 (b) source processes. The shaded areas denote how single PID atoms in (a) or groups of atoms in (b) are associated to unique (pink, green, orange), synergistic (blue), and redundant (red) components, providing a coarse-grained decomposition of the full GC. Red circles identify the atoms composing the bivariate GC $F_{X_1\rightarrow Y}$ in the two cases.}
    \label{fig_intro}
\end{figure}

In this section we e exploit the framework of partial information decomposition (PID) \cite{williams2010nonnegative} to dissect the full GC defined in (\ref{GCtime}) into components explaining how the different drivers $\{X_1,\ldots,X_N\}$ contribute to the overall information transferred towards the target process $Y$.
%The full GC defined in (\ref{GCtime}) can be interpreted as a measure of the overall information transferred collectively by the driver processes $\{X_1,\ldots,X_N\}$ to the target process $Y$. Here, we exploit the framework of partial information decomposition (PID) \cite{williams2010nonnegative} to dissect the full GC into components explaining how the different drivers contribute to the  overall information transfer towards the target;
Originally defined for multivariate random variables, PID establishes a mathematical framework designed to decompose the information shared by a target and several source variables into "atoms" quantifying the contribution brought to the target by specific collections of sources. Here, following the mathematical foundations of PID we decompose the full GC as
\begin{equation}
    F_{X\rightarrow Y} = \sum_{\alpha \in \mathcal{A}} F^{\delta}_{X_{\alpha} \rightarrow Y},
    \label{GCPID}
\end{equation}
where $\mathcal{A}$ collects all subsets of sources such that no subset is a superset of any other (e.g., $\mathcal{A}=\{\{1\}\{2\},\{1\},\{2\},\{12\}\}$ for $N=2$ driver processes, see Fig. \ref{fig_intro}a), and where $F^{\delta}_{X_{\alpha} \rightarrow Y}$ identifies the GC atom quantifying the information transferred from $X_{\alpha}$ to $Y$; for a given atom $\alpha=\{\alpha_1,\ldots,\alpha_J\}$, $X_{\alpha}$ is a collection of $J$ subsets of driver processes, where the subset $X_{\alpha_j}$ contains $N_j$ processes, $j=1,\ldots,J$. For instance, the atom $\alpha=\{\{2\},\{13\}\}$ in Fig. \ref{fig_intro}b identifies the $J=2$ subsets of drivers $X_{\alpha_1}=X_2$ and $X_{\alpha_2}=X_{13}=\{X_1,X_3\}$, and the corresponding GC quantifies the overlapping information transferred to $Y$ by $X_2$ and $\{X_1,X_3\}$.

%The collection of subsets of driver processes $X_{\alpha}$ is indexed by the atom $\alpha=\{\alpha_1,\ldots,\alpha_J\}$ whose subsets contain one or more drivers taken together, i.e. $X_{\alpha_j}$, $j=1,\ldots,J$;
%with $\alpha_j=\{j_1\cdots j_{N_j}\}\subset \{1,\ldots,N\}$;
%for instance, the atom $\alpha=\{\{2\},\{13\}\}$ identifies the $J=2$ subsets of drivers $X_{\alpha_1}=X_2$ and $X_{\alpha_2}=X_{13}=\{X_1,X_3\}$.% ($N_1=1, N_2=2$; Fig. \ref{fig_intro}b).

To solve the PID applied to the full GC it is necessary to provide a set of equations leading to derive the GC atoms on the r.h.s. of (\ref{GCPID}). To this end, following the spirit of PID \cite{williams2010nonnegative}, we introduce a so-called \textit{redundant GC} function which takes values over the lattice and satisfies the relation:
\begin{equation}
    F^{\cap}_{X_{\alpha} \rightarrow Y}=\sum_{\beta \preceq \alpha} F^{\delta}_{X_{\beta} \rightarrow Y},
    \label{RedundantGC}
\end{equation}
for each atom $\alpha \in \mathcal{A}$, where $\preceq$ identifies precedence based on the partial ordering imposed by the lattice structure (see Fig. 1; the ordering is from top to bottom). Moreover, according to the consistency equations of PID\cite{gutknecht2021bits}, the redundancy function computed for an atom formed by a single subset of sources is the bivariate GC from the sources to the target, i.e. $F^{\cap}_{X_{\alpha} \rightarrow Y}=F_{X_{\alpha} \rightarrow Y}$ if $J=1$. This implies that, according to (\ref{RedundantGC}), any bivariate GC is decomposed as the sum of GC atoms located at the same level or below in the lattice; for instance, the atoms of the GC $F_{X_1\rightarrow Y}$ are identified by the red circles in Fig. 1.

Given (\ref{RedundantGC}), if the redundant GC is known the GC amounts associated to all atoms can be retrieved iteratively starting from the bottom of the lattice or compactly via  M\"{o}bius inversion \cite{jansma2025fast}.
Therefore, solving the PID amounts to define the redundant GC $F^{\cap}_{X_{\alpha} \rightarrow Y}$ for each atom $\alpha \in \mathcal{A}$. Here, we follow the rationale of the recently proposed partial information rate decomposition \cite{faes2025partial}.
\textcolor{black}{We exploit the minimum mutual information principle at the pointwise level in the spectral domain to provide a frequency-specific definition of redundant GC which, if integrated over all frequencies, yields the time-domain redundant GC. Specifically, employing} frequency-domain expansion we define the redundant GC as the full-frequency integral of the pointwise \textit{spectral redundant GC}:
\begin{equation}
    F^{\cap}_{X_{\alpha} \rightarrow Y}:=\frac{1}{\pi}\int_{0}^{\pi}f^{\cap}_{X_{\alpha} \rightarrow Y}(\omega) \mathrm{d}\omega,
    \label{RedGCIntegral}
\end{equation}
which is in turn defined, at each normalized angular frequency $\omega \in [0,\pi]$, by taking the minimum of the spectral GCs directed to the target process from each subset of drivers composing the analyzed atom:
\begin{equation}
    f^{\cap}_{X_{\alpha} \rightarrow Y}(\omega):= \min_{j=1,\ldots,J} f_{X_{\alpha_j}\rightarrow Y}(\omega),
    \label{SpectralRedGC}
\end{equation}
where $f_{X_{\alpha_j}\rightarrow Y}(\omega)$ is the spectral GC defined in (\ref{SGC}).
Then, given the spectral redundant GCs computed via (\ref{SpectralRedGC}) over the lattice, applying M\"{o}bius inversion leads to derive the so-called \textit{spectral atomic GCs} expressed, for each atom $\alpha \in \mathcal{A}$, as
\begin{equation}
    f^{\delta}_{X_{\alpha} \rightarrow Y}(\omega)=f^{\cap}_{X_{\alpha} \rightarrow Y}(\omega)-\sum_{\beta \prec \alpha} f^{\delta}_{X_{\beta} \rightarrow Y}(\omega),
    \label{AtomicGC}
\end{equation}
which achieve a PID decomposition of the full GC analogous to (\ref{GCPID}) but expressed in the frequency domain:
\begin{equation}
    f_{X\rightarrow Y}(\omega) = \sum_{\alpha \in \mathcal{A}} f^{\delta}_{X_{\alpha} \rightarrow Y}(\omega).
    \label{GCPIDspect}
\end{equation}
\textcolor{black}{The main peculiarity of this approach for defining the redundant GC lies in its frequency-specific nature. Indeed, the fine-grained decomposition resulting in (\ref{GCPIDspect}) allows} dissecting the overall information brought by iso-frequency oscillations in the the driver processes to the oscillation occurring in the target at the same frequency 
into amounts which highlight a variety of interaction modes. For instance, the spectral atomic GCs relevant to the atoms $\{1\}$, $\{\{1\},\{2\}\}$, and $\{12\}$ refer respectively to the \textit{unique} information that $X_1$ provides exclusively to $Y$, to the \textit{redundant} (overlapping) information equally provided to $Y$ by $X_1$ and $X_2$, and to the \textit{synergistic} information provided to $Y$  by $X_1$ and $X_2$ taken jointly but not in isolation.
While for $N=2$ sources the unique, redundant and synergistic contributions are given by single atoms of the lattice, in the presence of more sources the number of atoms grows super-exponentially \cite{gutknecht2021bits}; (e.g. $|\mathcal{A}|=18$ for $N$=3, Fig. \ref{fig_intro}b). Therefore, to aid interpretability, it is appropriate to sum the information provided by multiple atoms yielding aggregate measures of unique, redundant and synergistic contributions of the sources to the target \cite{sparacino2025quantifying}. Here, the GC atoms are aggregated into $N+2$ quantities satisfying the coarse-grained decomposition of the full spectral GC:
\begin{equation}
    f_{X\rightarrow Y}(\omega) = f^\mathcal{R}_{X\rightarrow Y}(\omega) + f^\mathcal{S}_{X\rightarrow Y}(\omega) + \sum_{i=1}^{N} f^\mathcal{U}_{X_i\rightarrow Y}(\omega),
    \label{spectralCoarseGrainedGC}
\end{equation}
where $f^\mathcal{R}_{X\rightarrow Y}(\omega)$ and $f^\mathcal{S}_{X\rightarrow Y}(\omega)$ are the \textit{redundant} and the \textit{synergistic} spectral GCs from all drivers to the target, and $f^\mathcal{U}_{X_i\rightarrow Y}(\omega)$ is the \textit{unique} spectral GC from the $i^{\mathrm{th}}$ driver to the target. In this work, coarse-graining is performed following the aggregation rules established in \cite{rosas2020reconciling}, \textcolor{black}{formalized in the Supplementary Material (sect. II.D) and} illustrated in Fig. 1b for the case of $N=3$ sources.

%Exploiting the fact that the backbone of the PID lattice can be used equivalently in the spectral or time domains to define frequency-specific or aggregate measures, the 
Crucially, the frequency-domain PID (\ref{GCPIDspect}) is closely related to the time-domain PID (\ref{GCPID}) through the spectral integration property, which is enabled by the redundancy definition given by (\ref{RedGCIntegral}) \cite{faes2025partial,sparacino2025quantifying}. In fact, performing whole-band integration of (\ref{AtomicGC}) and using (\ref{RedGCIntegral}) recovers (\ref{RedundantGC}), which shows that the time-domain atomic GC is the full-frequency integral of the spectral atomic GC, i.e. $F^{\delta}_{X_{\alpha} \rightarrow Y}=\frac{1}{\pi}\int_0^{\pi}f^{\delta}_{X_{\alpha} \rightarrow Y}(\omega)\mathrm{d}\omega$.
In the same way, whole-band integration of (\ref{spectralCoarseGrainedGC}) yields a coarse-grained decomposition of the full GC in the time domain evidencing redundant, synergistic and unique components:
\begin{equation}
    F_{X\rightarrow Y} = F^\mathcal{R}_{X\rightarrow Y} + F^\mathcal{S}_{X\rightarrow Y} + \sum_{i=1}^{N} F^\mathcal{U}_{X_i\rightarrow Y}.
    \label{CoarseGrainedGC}
\end{equation}
%Although the PID  is defined in the spectral domain, a corresponding time-domain PID of the full GC can be obtained  by exploiting the redundancy definition given in (\ref{RedGCIntegral}) and performing full-frequency integration of (\ref{AtomicGC}). The resulting atomic GC, $F^\delta_{X_{\alpha}\rightarrow Y}$, can then be aggregated using the same rules exploited in (\ref{spectralCoarseGrainedGC}) to obtain the time-domain of redundant, synergistic and unique GCs which satisfy: 
%Crucially, the components of the decomposition (\ref{CoarseGrainedGC}) could be obtained directly from full frequency integration of (\ref{spectralCoarseGrainedGC}).
Thus, there exists a tight relation between the fine- and coarse-grained time-domain decompositions of the full GC stated in (\ref{GCPID}) and (\ref{CoarseGrainedGC}) and their spectral expansions stated in (\ref{GCPIDspect}) and (\ref{spectralCoarseGrainedGC}).
This relation allows also to tailor the GC decompositions to predetermined frequency bands with practical meaning: limiting the integral to the frequency range $[\omega_1, \omega_2]$ in (\ref{RedGCIntegral}) allows to focus on the portion of information transferred from the drivers to the target only relevant to the oscillatory components with frequencies included between $\omega_1$ and $\omega_2$; this possibility will be exploited in Sect. \ref{ssec:DataAnalysis} to perform the analysis with regard to specific physiological oscillations.

Notably, the fine- and coarse-grained decompositions presented in this Section allow not only to dissect the full GC into atoms of information flow, but also to provide a meaningful interpretation for the existing formulations of the GC. In particular, the application of (\ref{RedundantGC}) to  $F_{X_i}\rightarrow Y$ shows that this bivariate GC is composed by unique and redundant atoms (see Fig. 1 for $i=1$); on the other hand it is easy to show that the conditional GC is composed by unique and synergistic atoms. These interpretations formalize and quantify previous intuitions based on the comparison of bivariate and conditional forms of the GC in simulated settings \cite{stramaglia2014synergy,stramaglia2016synergetic}.

\subsection{Computation based on State-Space Models} \label{sec:computation}
A main issue with the computation of GC from VAR models is the fact that the order of the reduced model (\ref{redVARmodel}) is theoretically infinite, which hampers the computation of GC estimates obtained by truncating the order to a finite value \cite{faes2017interpretability}. In this work, we overcome this issue by shifting from the VAR representation to the state space (SS) representation of multivariate processes \cite{barnett2015granger}. As the class of SS models is closed under the formation of reduced models, its use allows to formalize GC without theoretical approximations, thus offering high computational reliability.

To represent the VAR model (\ref{VARmodel}) as an SS model, a so-called state process $\bar{S}=\{\bar{S}_n\}_{n\in \mathbb{Z}}$ is defined by concatenating the past states of the observed process $S$ as $\bar{S}_n=[S_{n-1}^\intercal\cdots S_{n-p}^\intercal]^\intercal$, and the two processes are related by
\begin{subequations}
\begin{align}
    S_n &= \textbf{C} \bar{S}_n+U_n \label{ObsEq} \\
    \bar{S}_{n+1} &= \textbf{A} \bar{S}_n + \textbf{K}U_n \label{StEq}
\end{align} \label{SSmodel}
\end{subequations}
\noindent{where (\ref{ObsEq}) is the observation equation linking, at each time step $n$, the observed and state processes $S_n$ and $\bar{S}_n$ through the observation matrix $\textbf{C}$, and (\ref{StEq}) is the state transition equation determining the one-step ahead update of the state process through the transition matrix $\textbf{A}$; $\textbf{K}$ is the Kalman gain matrix relating the observation error $U_n$ to its representation in the state equation.}
Then, as for the VAR representation, to compute the GC it is necessary to identify a restricted SS model for the $K$-dimensional sub-process $Z=\{X_{\alpha_j},Y\}$ where $X_{\alpha_j}\subset X$ includes $N_j=K-1$ drivers. Such model can be written in the form \cite{barnett2015granger,faes2022new}:
\begin{subequations}
\begin{align}
    Z_n &= \tilde{\textbf{C}} \bar{S}_n+W_n, \label{ObsEq_restr} \\
    \bar{S}_{n+1} &= \tilde{\textbf{A}} \bar{S}_n + \tilde{\textbf{K}}W_n, \label{StEq_restr}
\end{align} \label{SSmodel_restr}
\end{subequations}
where the parameters are the matrices $\tilde{\textbf{A}}$, $\tilde{\textbf{C}}$, $\tilde{\textbf{K}}$, and $\boldsymbol{\Sigma}_W$, \textcolor{black}{which are derived from the parameters of the full SS model (\ref{SSmodel}) as described in the Supplementary Material (Sect. S.II).}

After identification, the model (\ref{SSmodel_restr}) is analyzed in the frequency domain taking the FT of (\ref{StEq_restr}) and combining it with the FT of (\ref{ObsEq_restr}) to obtain $Z(\omega)=\textbf{H}(\omega)W(\omega)$, which evidences the $K\times K$ TF matrix:
\begin{equation}
    \textbf{H}(\omega)=\textbf{I}_K+\tilde{\textbf{C}}[\textbf{I}_{Mp}-\tilde{\textbf{A}}e^{-\textbf{j}\omega}]^{-1}+\tilde{\textbf{K}}e^{-\textbf{j}\omega},
    \label{TF}
\end{equation}
where $\textbf{j}=\sqrt{-1}$. The TF matrix $\textbf{H}(\omega)$ in (\ref{TF}) is equivalent to the one describing the VAR model (\ref{redVARmodel},\ref{redVARdec}), while it differs in the last column from the TF matrix $\tilde{\textbf{H}}(\omega)$ of the VAR model with instantaneous effects formed by (\ref{redVAR2tilda}) and (\ref{redVAR1}). Nevertheless, the fact that the two TF matrices share the first $N_j$ columns (i.e., $H_{11}(\omega)=\tilde{H}_{11}(\omega)$, $H_{21}(\omega)=\tilde{H}_{21}(\omega)$) allows combining (\ref{PSDY}) and (\ref{SGC}) to compute the spectral GC from $X_{\alpha_j}$ to $Y$ as:
\begin{equation}
    f_{X_{\alpha_j}\rightarrow Y}(\omega)=\ln \frac{P_Y(\omega)}{P_Y(\omega)-H_{21}(\omega)\tilde{\Sigma}_{1}H^*_{21}(\omega)}.
    \label{SGC_SS}
\end{equation}
The spectral GC (\ref{SGC_SS}), computed at varying the processes included in $X_{\alpha_j}$ and used as in (\ref{SpectralRedGC}), constitutes the building block of the decomposition of the full GC (\ref{GCtime}) into Granger-causal atoms of information transfer.

%in the next subsection, the "reduced" GCs obtained as in (\ref{GCtime_sub}) at varying the processes included in $X_{\alpha_j}$ will be used as the building blocks of the decomposition of the "full" GC (\ref{GCtime}) into Granger-causal atoms of information transfer.

\section{Validation on Simulations} \label{sec:simulations}

\textcolor{black}{In this section, the PDGC framework is illustrated using simulated VAR models, for which all measures can be evaluated under controlled ground-truth conditions. The simulations were designed to: (i) demonstrate how the theoretical PDGC profiles computed from the known model parameters capture unique, redundant, and synergistic information transfer associated with the oscillatory and causal structure of the network; and (ii) assess the reliability of the corresponding estimates obtained from finite-length realizations.}

\textcolor{black}{We first designed a Gaussian system with $M=4$ processes to reproduce redundant and synergistic causal interactions from the drivers $X=[X_1,X_2,X_3]$ to the target $Y$. Stochastic oscillations were imposed at frequencies $\bar{f}_1$=0.1 Hz, $\bar{f}_2$=0.3 Hz, and $\bar{f}_3$=0.4 Hz for the three drivers (simulated sampling frequency $\bar{f}_s$=1 Hz). The model was configured so that $X_2$ and $X_3$ exert direct causal influences on $Y$, whereas $X_1$ affects $Y$ only indirectly through the mediator $X_2$ (Fig. \ref{fig_rev}a; equations and parameter values are reported in the Supplemental Material, Sect. III).}
\textcolor{black}{Applying PDGC to the known VAR parameters yields the theoretical spectral functions shown in Fig. \ref{fig_rev}a-c. The PSDs of the four processes exhibit distinct peaks at the imposed oscillation frequencies (Fig. \ref{fig_rev}a), while the full spectral GC reflects the frequency-specific information transfer from the three drivers to the target (Fig. \ref{fig_rev}b). Computing the spectral GC for all driver subsets (Eq. 9), deriving the spectral redundant GCs (Eq. 13), and performing the frequency-specific M\"{o}bius inversion (Eq. 14) yields the 18 spectral atomic GCs displayed on the PID lattice in Fig. \ref{fig_rev}c.
These atoms decompose the overall information transfer into meaningful elementary components. Specifically, $f^{\delta}_{X_{\{2\}}\to Y}$ and $f^{\delta}_{X_{\{3\}\to Y}}$ exhibit distinct peaks at $\bar{f}_2$ and $\bar{f}_3$, identifying unique information transfer associated with the direct causal effects of $X_2$ and $X_3$. The atomic GC $f^{\delta}_{X_{\{1\}\{2\}}\to Y}$ peaks at $\bar{f}_1$, revealing redundant information transfer generated by the indirect pathway from $X_1$ to $Y$ through $X_2$. The atomic GC $f^{\delta}_{X_{\{23\}}\to Y}$ peaks between $\bar{f}_2$ and $\bar{f}_3$, identifying synergistic information transfer that emerges only when $X_2$ and $X_3$ are considered jointly.
Since the remaining atomic GCs are null or negligible, these components largely determine the coarse-grained spectral GC profiles obtained by grouping the atoms according to the coarse-graining rules (Fig. \ref{fig_rev}d). Accordingly, the coarse-grained GCs identify the unique information transfer associated with the direct pathways $X_2\to Y$ (green) and $X_3\to Y$ (orange), the redundant transfer generated by the cascade $X_1\to X_2 \to Y$ (red), and the synergistic transfer associated with the collider $X_2\rightarrow Y \leftarrow X_3$ (blue).}

\begin{figure*} [t!]
    \centering
    \includegraphics[scale= 1.25]{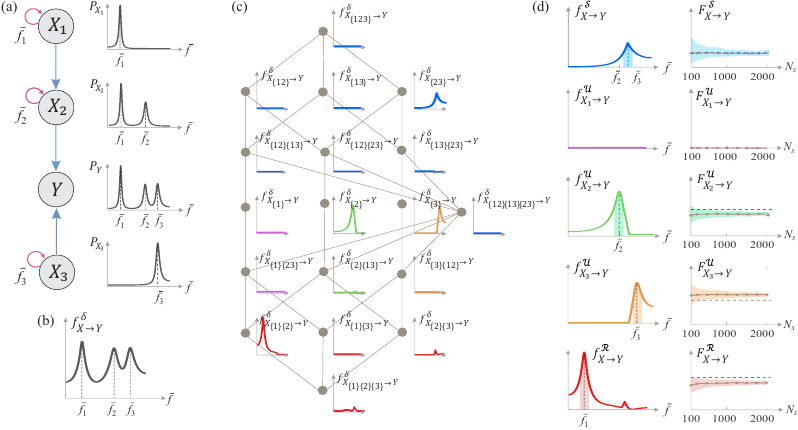}
    \caption{\textcolor{black}{PDGC analysis in a simulated VAR system with three drivers and a target process. (a) Connectivity structure and power spectral densities of the four simulated processes. (b) Spectral GC from the drivers to the target. (c) Spectral atomic GCs which compose the full GC according to (\ref{GCPIDspect}), depicted over the GC redundancy lattice. (d) Theoretical spectral profiles (left) and band-limited time-domain estimates from finite-length realizations (right) of the synergistic (blue), unique (pink, green orange) and redundant (red) coarse-grained components of the full GC.}}
    \label{fig_rev}
\end{figure*}

\textcolor{black}{The performance of practical estimation was assessed by generating 100 realizations with lengths ranging from 100 to 2000 samples and computing the coarse-grained PDGC functions from the SS parameters estimated for each realization (VAR identification: ordinary least squares; model order selection: Bayesian Information Criterion \cite{faes2012measuring}). To summarize estimation performance, both theoretical and estimated spectral profiles were integrated over a 0.05 Hz frequency band centered at the peak of each PDGC component (Fig. \ref{fig_rev}d). The results show the expected reduction in estimation variance as the number of samples increases, with the width of the $5^{\mathrm{th}}$-$95^{\mathrm{th}}$ percentile interval decreasing with $N_s$, together with an asymptotic bias affecting the redundant and unique components. This behavior is consistent with previous findings showing that the accuracy of parametric GC estimation critically depends on the ratio between the available data points and the number of model parameters \cite{antonacci2024measuring}. Additional simulations reported in the Supplementary material, performed at varying the coupling parameters and the VAR model order while keeping the data length fixed ($N_s=250$), confirmed that the estimated GC components closely follow the theoretical profiles despite the bias (Fig. S2a)} \textcolor{black}{and are relatively robust to  model order misspecification (Fig. S2b).}

\textcolor{black}{A second simulation, reported in the Supplemental Material (Sect. III), compares the proposed PDGC measures with the corresponding PIRD measures \cite{faes2025predictive} in a Gaussian system with $M=3$ processes analyzed by varying the target process. The comparison shows that PIRD reduces to PDGC when all causal interactions are directed exclusively toward the selected target, whereas PDGC additionally reveals high-order Granger-causal effects arising from interaction patterns that include causal influences originating from the target (Fig. S3).}

\section{Application to Cardiovascular and Cerebrovascular Networks} \label{sec:application}
To illustrate the potential of the proposed framework for characterizing the human homeostatic regulation across different physio-pathological states, we analyzed the multivariate interactions underlying cardiovascular (CV) and cerebrovascular (CB) control in a group of patients prone to develop orthostatic syncope compared with healthy controls. In this context, several studies have investigated CV and CB interactions from the spontaneous variability of heart rate, cerebral blood flow velocity, arterial pressure and respiration \cite{mullen1997system,ocon2009increased,medow2014altered, gelpi2022dynamic,pernice2022spectral,sparacino2023method,porta2023concomitant}. Here, we propose the high-order decomposition of the spectral GC as a tool to elicit the intertwined physiological mechanisms underlying the short-term control of heart rate and cerebral blood flow at rest and during postural stress.

\subsection{Experimental Protocol and Time series extraction} \label{ssec:Protocol}
We analyze data previously collected for the description of cardiovascular and cerebrovascular determinants of neurally mediated syncope \cite{gelpi2022dynamic,pernice2022spectral}.
The study involved thirteen subjects prone to syncope (SYNC, 5 males; age: 28 $\pm$ 9 yrs) and thirteen age and gender-matched control individuals (nonSYNC, 5 males; age: 27 $\pm$ 8 years) enrolled at the Neurology Division of Sacro Cuore Hospital, Negrar, Italy. Participants were evaluated at rest in the supine position (REST, 10 min) and during head-up tilt (HUT, up to 40 min; 60° degree of tilt). All SYNC participants developed pre-syncope during the HUT phase, while none of the nonSYNC participants experienced noticeable symptoms for the whole duration of the experiment. All the participants signed a written informed consent to join the experimental study approved by the local Ethical Committee \cite{pernice2022spectral}.  

The recorded signals were the electrocardiogram (ECG, lead II), the finger  continuous arterial blood pressure (ABP) measured by the photoplethysmographic volume-clamp method, the cerebral blood velocity (CBV) signal measured from the middle cerebral artery through a transcranial Doppler ultrasound device, and the respiratory amplitude signal through a piezoelectric thoracic belt. From these signals synchronously recorded  at a sampling rate of 1 kHz, the following time series were extracted:
heart period (HP), measuring the temporal distance between two consecutive R-waves peaks of the ECG signal; the systolic arterial pressure (SAP), measured as the maximum of the ABP signal taken inside each detected HP; the mean arterial pressure (MAP) and the mean CBV (MCBV), measured as the integral of the ABP and CBV waveforms taken within each detected diastolic pulse interval and divided by the duration of the interval itself; and respiration (RESP) obtained by sampling the respiratory signal in correspondence of each detected R-peak. After removing slow trends (zero-phase AR low-pass filter, cutoff frequency 0.0156 cycles/beat), synchronous stationary zero-mean sequences of 250 points were extracted for the five time series \cite{pernice2022spectral}.

\subsection{Data and statistical analysis} \label{ssec:DataAnalysis}
The pipeline adopted to perform PDGC analysis of CV and CB interactions for each SYNC and nonSYNC participant in the REST and HUT conditions is described in detail in the following, and illustrated for a representative participant in {Fig. \ref{fig_example}. 
First, the time series obtained as described in Sect. \ref{ssec:Protocol} were taken as realizations of five discrete-time stationary random processes describing the dynamics of RESP, SAP, MAP, HP, and MCBV (respectively, processes $R$,$S$,$M$,$H$,$V$; Fig. \ref{fig_example}a). 
These processes are typically studied in the frequency domain by considering them as sampled uniformly with a sampling period equal to the mean HP, i.e. with sampling frequency $\bar{f}_s=\frac{1}{\mathbb{E}[H_n]}$; accordingly, all spectral functions can be represented as a function of the frequency $\bar{f}$ varying in the range $[0, \frac{\bar{f}_s}{2}]$ (e.g., the PSD is reported in Fig. \ref{fig_example}b).

\begin{figure*} [t!]
    \centering
    \includegraphics[scale= 1]{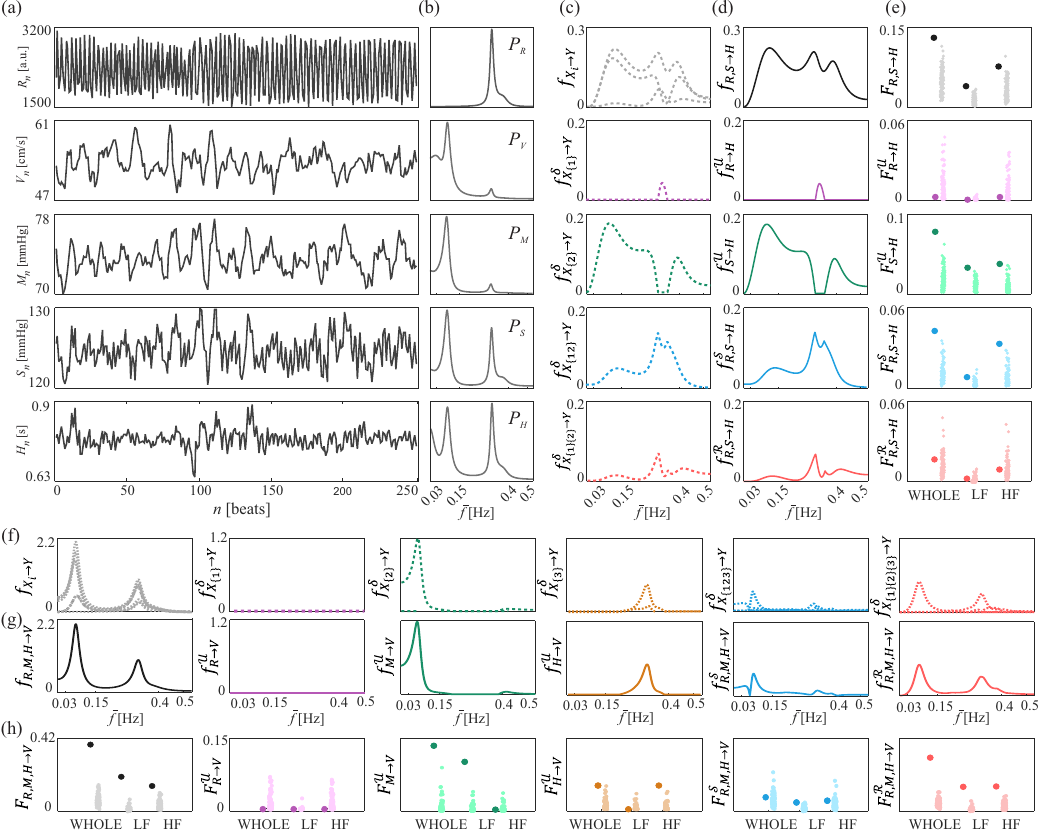}
    \caption{Example of PDGC analyses performed for a representative SYNC patient in the resting condition. Panels depict the analyzed variability series (a) together with their estimated power spectrum (b), and the PDGC measures computed to analyze CV interactions (c-e; target $Y=H$, drivers $X=\{R,S\}$) and CB interactions (f-h; target $Y=V$, drivers $X=\{R,M,H\}$). In the two settings, panels depict the spectral and atomic GCs (c,f), the full spectral GC and the terms of its coarse-grained decomposition (d,g), and the time-domain values of the full and coarse-grained GCs computed along with their surrogate distributions (e,h).}
    \label{fig_example}
\end{figure*}

We considered two settings typically adopted for investigating the short-term control of heart rate and cerebral blood flow: to study CV interactions (Fig. \ref{fig_example}c-e), we take HP as the target process and RESP and SAP as driver processes (i.e., $Y=H$, $X=\{R,S\}$);
%the overall vector process $S=\{R,S,H\}$ was analyzed considering HP as the target process and RESP and SAP as driver processes (i.e., $Y=H$, $X=\{R,S\}$);
to study CB interactions (Fig. \ref{fig_example}f-h), we take MCBV as the target process and RESP, MAP and HP as driver processes (i.e., $Y=V$, $X=\{R,M,H\}$).
In each setting, a VAR model was identified on the selected time series (ordinary least square regression, model order $p$ set in the range 3-12 via the Bayesian Information Criterion \cite{faes2012measuring}).
The estimated VAR parameters $\hat{\textbf{A}}_k$ and $\hat{\boldsymbol{\Sigma}}_U$ were then used to compute the corresponding SS parameters, from which the spectral GC was obtained as in Sect. \ref{sec:computation} (full GC: Fig. \ref{fig_example}d,g, black). 
The analysis was iterated over the atoms of the GC lattice as in Sect. \ref{sec:HOGC}, computing for each atom the spectral GCs  of every component of the atom (Fig. \ref{fig_example}c,f, gray), deriving the spectral redundant GC of the atom according to (\ref{SpectralRedGC}) and applying M\"{o}bius inversion to get the spectral atomic GCs (Fig. \ref{fig_example}c,f, dashed-colored). These GCs were then aggregated to obtain the unique, synergistic and redundant spectral GCs (respectively, solid pink/green/orange, solid blue and solid red in Fig. \ref{fig_example}d,g).
Finally, time-domain values for the full GC and for its coarse-grained unique, synergistic and redundant components were obtained by integrating the spectral GCs  over the whole frequency range ($\bar{f}\in [0,\frac{\bar{f}_s}{2}]$), and also within specific frequency bands of physiological interest, i.e. the low-frequency band (LF, $\bar{f}\in $ [0.03 0.15] Hz) and the high-frequency band (HF, $\bar{f}\in$ [0.15 0.4] Hz) (Fig. \ref{fig_example}e,h, big circles).

To assess the statistical significance of the time-domain GCs on a single-subject basis, an approach based on surrogate data was employed. Specifically, the iterative amplitude-adjusted FT (IAAFT) procedure \cite{schreiber1996improved} was applied separately for each time series, yielding a set of surrogate series with preserved amplitude distribution and PSD but destroyed cross-PSD; one hundred sets of surrogate time series were generated in this way, and estimation of all GCs was repeated for each surrogate set
(Fig. \ref{fig_example}e,h, small circles). Then, a significance test based on percentiles was run with 5$\%$ significance: any considered GC measure was deemed as statistically significant if its estimate obtained for the original series was above the $95^{\mathrm{th}}$ percentile of the distribution derived from the surrogates. For instance in the example of Fig. \ref{fig_example}e, the full and unique GCs $F_{S,R\rightarrow H}$ and $F^{\mathcal{U}}_{S\rightarrow H}$ are deemed as statistically significant in all bands, the redundant and unique GCs $F^{\mathcal{R}}_{S,R\rightarrow H}$ and $F^{\mathcal{U}}_{R\rightarrow H}$ are deemed as non significant, and the synergistic GC $F^{\mathcal{S}}_{S,R\rightarrow H}$ is deemed as significant in the full and HF bands but not in the LF band. 

The analyses described above were repeated for the series measured from all SYNC ad nonSYNC participants in both REST and HUT conditions, obtaining estimates of the full GC and of its coarse-grained components over all frequencies and within the LF and HF bands, as well as their statistical significance.  
At the group level, statistical analysis was performed comparing, for each measure, the distributions across participants obtained at REST and during HUT by means of the non-parametric Wilcoxon signed-rank test for paired data performed with $5\%$ significance.

\begin{figure*}[t]
    \centering
    \includegraphics[width=\linewidth]{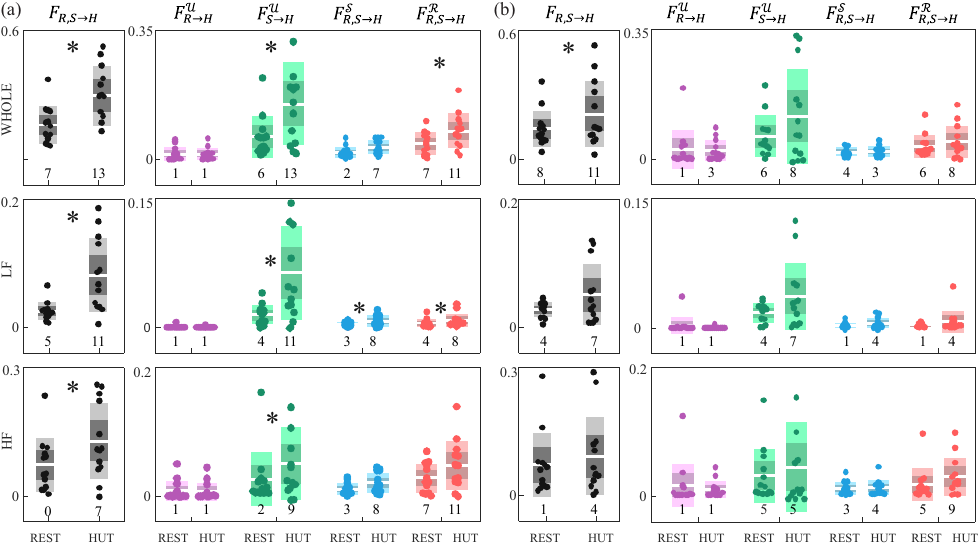}
    \caption{PDGC analysis of CV interactions in nonSYNC subjects (a) and SYNC patients (b). Panels depict the boxplots and individual values of the of the full GC from $\{$RESP,SAP$\}$ to HP (gray), as well as of its coarse grained decomposition evidencing the unique GCs originating from RESP (violet) and SAP (green), the synergistic GC (blue) and the redundant GC (red), computed at REST and during HUT through whole-band integration (top panels) or through integration within the LF band (middle panels) or the HF band (bottom panels); the number of nonSYNC or SYNC participants (out of 13) for whom each measure was deemed as statistically significant according to surrogate data analysis is also reported in each panel. *, $p<0.05$ REST vs. HUT, Wilcoxon signed-rank test.}
    \label{fig3}
\end{figure*}

\subsection{Results and Discussion} \label{subsec:results}
\subsubsection{PDGC Analysis of Cardiovascular Interactions}
Fig. \ref{fig3} reports the results of the PDGC analysis of CV interactions, performed by decomposing the Granger-causal effects of RESP and SAP on HP in both nonSYNC and SYNC groups and considering both whole-band and band-specific measures.
The overall predictive information transferred from RESP and SAP to HP (Fig 3a,b, left panels) quantified by the full GC was statistically significant in the majority of participants if assessed through whole-band integration, and in a lower number of participants if assessed in the LF band and especially in the HF band; this number was higher in HUT than in REST, with a more marked increase in nonSYNC subjects compared to SYNC patients. These results about the detection of multivariate GC were confirmed by the analysis of the amount of information transferred from RESP and SAP to HP; the full GC resulted higher during HUT than at REST, with an increase that was marked and statistically significant in all bands for the nonSYNC subjects and was limited to the whole-band measure for the SYNC patients.
%Overall, these findings reflect the response of the cardiovascular system to orthostatic stress, involving an enhancement of CV interactions that was evident especially in the healthy subjects and somewhat blunted in the individuals developing postural syncope.

The different response to HUT displayed by SYNC and nonSYNC participants becomes more striking by looking at the coarse-grained decomposition of the full GC (Fig. 3a,b, right panels). In fact, while the nonSYNC group exhibited a marked and statistically significant increase of the unique GC SAP$\rightarrow$HP in all bands, of the redundant GC $\{$RESP,SAP$\}\rightarrow$ HP in the full and LF bands, and of the synergistic GC $\{$RESP,SAP$\}\rightarrow$ HP in the LF band while moving from REST to HUT, no significant difference in any of the components of the full GC was found in the SYNC group. These results were mirrored by the number of participants displaying statistically significant GC components.
%Overall, the high-order causal analysis suggests that the physiological response to postural stress is reflected by increased effects of arterial pressure on heart rate variability, both unique and redundant with respiration, which are blunted in the presence of orthostatic intolerance leading to syncope development. 

Physiologically, it is well-known that postural stress elicits a shift in the sympatho-vagal balance towards sympathetic activation and parasympathetic deactivation, which is associated with an engagement of the baroreflex mechanism and a reduction of respiratory sinus arrhythmia \cite{cohen2002short}. Several methodological studies have documented that this physiological response is reflected by an increase with tilt of the Granger-causal effects from SAP to HP computed either in a bivariate fashion or conditioning on RESP, and a concomitant decrease of the causal influences of RESP on HP \cite{faes2013mechanisms,faes2014lag,faes2015information}. These findings were largely confirmed by the application of our approach for causal analysis to the healthy control group. On the other hand, the blunted or absent variation with tilt of the GC measures computed for the syncope patients documents an ineffective response to postural stress, in agreement with previous studies that identified baroreflex dysfunction reflected by a drop of causality RR$\rightarrow$SAP as a marker of the onset of presyncope \cite{faes2005causal,faes2013mechanisms}.
Besides confirming the literature, the PDGC analysis led us to refine previous findings identifying the information atoms related to stress-induced variations of the GC measures. Specifically, the increased information transfer from SAP to HP during tilt is ascribed to both the unique and the redundant components of the GC, which can be associated respectively to pure (respiration-unrelated) baroreflex effects and to influences driven by respiration. Moreover, the presence of an important redundant component, together with the low or negligible unique GC detected from RESP to HP, supports the interpretation of RSA as a mechanism due to HF respiratory-induced blood pressure oscillations that are translated into heart rate oscillations by the baroreflex \cite{karemaker2009counterpoint}.

\subsubsection{PDGC Analysis of Cerebrovascular Interactions}
Fig. 4 reports the results of the PDGC analysis of CB interactions, performed by decomposing the Granger-causal effects of RESP, MAP and HP on MCBV in nonSYNC and SYNC groups and considering whole-band and band-specific measures.
The full GC (Fig. 4a,b, left panels) was high and statistically significant in both groups and conditions particularly when computed through full- or LF-band integration of the spectral functions, documenting a significant dynamic transfer of information from the cardiovascular variables to cerebral blood flow.
Nevertheless, while this overall predictive information did not exhibit statistically significant changes in the transition from REST to HUT in the non-SYNC group, it increased significantly during HUT in the SYNC group; in the latter, the overall time-domain increase is ascribed to the LF band of the spectrum, while the HF contribution was unaffected. %These results indicate the increased overall information transfer from the cardiovascular processes to the cerebral blood velocity process in response to postural stress as a marker of orthostatic intolerance.

\begin{figure*}[t]
    \centering
    \includegraphics[width=\linewidth]{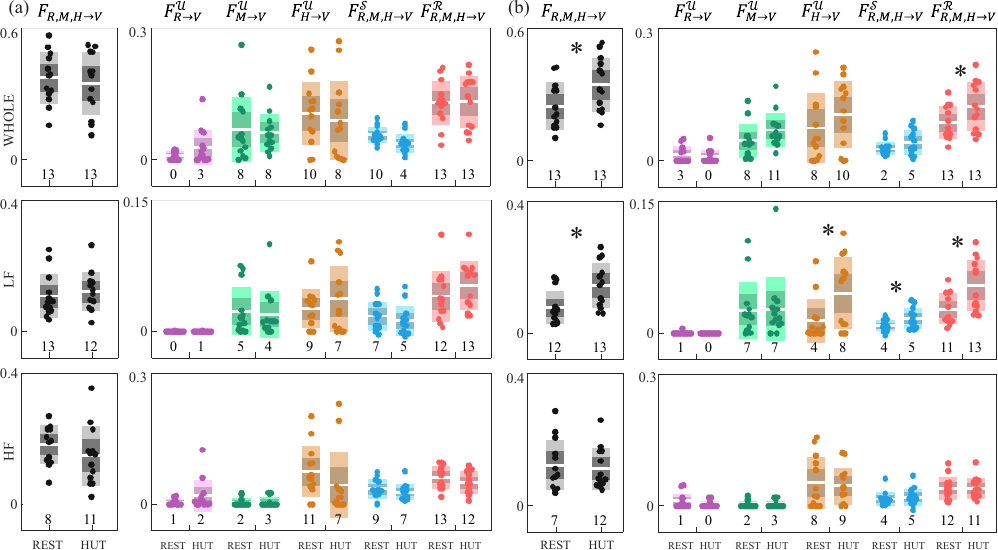}
    \caption{PDGC analysis of CB interactions in nonSYNC subjects (a) and SYNC patients (b). Panels depict the boxplots and individual values of the full GC from $\{$RESP,MAP,HP$\}$ to MCBV (gray), as well as of its coarse grained decomposition evidencing the unique GCs originating from RESP (violet), MAP (green) and HP (orange), the synergistic GC (blue) and the redundant GC (red), computed both at REST and during HUT through whole-band integration (top panels) or through integration within the LF band (middle panels) or the HF band (bottom panels); the number of nonSYNC or SYNC participants (out of 13) for whom each measure was deemed as statistically significant according to surrogate data analysis is also reported in each panel. *, $p<0.05$ REST vs. HUT, Wilcoxon signed-rank test.}
    \label{fig4}
\end{figure*}

The above results were further elucidated by dissecting the full GC into the coarse-grained unique GCs along the causal directions RESP$\rightarrow$MCBV, MAP$\rightarrow$MCBV and HP$\rightarrow$MCBV, and the coarse-grained redundant and synergistic GCs from \{RESP, MAP, HP\} to MCBV (Fig. 4a,b, right panels).
We found that, in the healthy controls, none of the components of the full GC was altered significantly moving from rest to HUT (Fig. 4a). On the contrary, in the syncope patients the transition from REST to HUT evidenced a general tendency to increase of the GC components, with statistically significant higher values observed during HUT for the redundant GC assessed via whole-band integration, and for the unique GC HP$\rightarrow$MCBV, redundant GC and synergistic GC assessed in the LF band; the increase of the time-domain unique GC MAP$\rightarrow$MCBV was close to statistical significance ($p=0.057$).
%These findings support the trends observed for the full GC, and indicate that both redundant and synergistic high-order interactions between LF oscillations of arterial pressure and heart rate determine the tilt-induced increase of the information transferred to cerebral blood velocity in syncope patients but not in healthy controls. Moreover, all analyses performed indicate a negligible role of respiration-related high-frequency oscillations in altering the information transfer observed during tilt in the syncope group.

These results confirm and complement from the broader point of view of multivariate high-order analysis previous findings in the literature documenting an increased causal coupling from arterial pressure to mean cerebral blood flow variables occurring during orthostatic stress in patients with poor postural tolerance \cite{ocon2009increased,medow2014altered,pernice2022spectral}. In these and other works an increased causal coupling during stress is interpreted as a marker of less effective dynamic cerebral autoregulation, the physiological mechanism whereby MCBV is maintained relatively constant in spite of changes of peripheral variables like MAP. The application of PDGC led us to generalize this result highlighting the role of other physiological variables, in particular HP variability. In fact, our results provide further explanation of the increased bivariate GC MAP$\rightarrow$MCBV previously detected in syncope patients during HUT \cite{pernice2022spectral}, showing that such an increase is due only in part to the unique contribution to MCBV provided by MAP, and is determined mainly by the HOIs between MAP and HP. More generally, since MAP and HP reciprocally affect each other through baroreflex feedback and mechanical feedforward mechanisms whose effectiveness is challenged by postural stress \cite{mullen1997system}, one may expect that the tilt-induced alterations in this closed-loop relation and in the direct effect that the cardiovascular variables have on MCBV are reflected in the enhanced redundant and synergistic contributions observed in the SYNC group. 
Notably, since the observed variations do not include the unique contribution of RESP to MCBV, and were not detected in any measure within the HF band that reflects respiration-related variability, we suggest that respiratory influences are not involved in the altered CB response to postural stress of syncope patients.

%\subsubsection{Commentary}
%Our findings elucidate the response of the CV and CB systems to orthostatic stress, revealing markedly different physiological adaptations in the healthy controls and in the patients prone to the development of orthostatic syncope. In particular, postural stress induced an enhancement of CV interactions related to increased causal effects of SAP on HP variability, which was evident in the nonSYNC subjects but blunted in the SYNC patients. On the contrary, CB interactions were modified in SYNC patients, but not in nonSYNC controls, via a tilt-induced increase of the causal effects of HP and AP on MCBV. While these opposite trends could be inferred from the multivariate GC computed at rest and during tilt, the high-order causal analysis evidenced them more strikingly and allowed to elicit physiological mechanisms related to specific components of the information transfer and/or to specific frequency bands, as discussed in the following. 
%. Moreover, the partial decomposition of the multivariate GC allowed to elicit  

 \section{Discussion}
The present work formalizes the partial information decomposition of Granger causality (PDGC) as a framework to quantify unique, redundant, and synergistic causal effects exerted by a set of source processes on an assigned target. A key feature of the proposed PDGC measures is their spectral formulation, which enables data-driven analysis of Granger-causal HOIs in networks of oscillatory processes, where a purely time-domain approach would inevitably mix contributions arising from different frequency bands.
\textcolor{black}{The need to operate in the frequency domain motivated the definition of the redundancy function chosen to operationalize PID in this setting. The choice of an appropriate redundancy function remains one of the main open issues in the PID literature, where several competing definitions have been proposed, each leading to different decompositions \cite{liardi2026mathematical}. The redundant GC introduced in Eqs. (\ref{RedGCIntegral},\ref{SpectralRedGC}) offers several advantages. First, it enables decomposition at individual frequencies, providing a perspective analogous to local PID approaches \cite{finn2018pointwise}, but tailored to iso-frequency components and therefore particularly suited to frequency-domain analysis. Moreover, unlike pointwise time-domain methods \cite{finn2018pointwise}, the proposed decomposition yields non-negative contributions at every frequency, ensuring straightforward interpretation of the resulting spectral causality components.
Moreover, integrating the pointwise redundant GC over predefined frequency bands naturally yields band-specific PID measures, while integration over the entire spectrum provides our time-domain definition of the redundant GC. Notably, this quantity is always smaller than the redundancy defined via the minimum-information criterion directly in the time domain \cite{sparacino2025decomposing}. This difference reflects the non-additivity of minimum-information-based redundancy across frequencies and, at the same time, provides a less conservative estimate that may mitigate the tendency of minimum-information definitions to overestimate redundancy \cite{liardi2026mathematical}.
More generally, this comparison highlights that PID results depend fundamentally on the adopted redundancy definition. Although frequency-domain formulations remain largely unexplored, alternative redundancy measures developed in the time domain could, in principle, be extended to the Granger-causality framework. In particular, accounting for causal interactions among the source processes may lead to more interpretable redundancy measures within the GC formalism. A systematic comparison of alternative redundancy definitions therefore represents an important direction for future research.}

\textcolor{black}{In both its conceptual formulation and practical implementation, PDGC belongs to a family of frameworks recently developed to quantify dynamic HOIs in networks of random processes \cite{faes2022new,faes2025partial}. The O-information rate (OIR, \cite{faes2022new}) and partial information rate decomposition (PIRD, \cite{faes2025partial}) frameworks can be regarded as predecessors of PDGC, as they provide spectral and corresponding time-domain measures of synergy and redundancy that extend to dynamic processes concepts originally developed for groups of random variables with static interactions \cite{rosas2019quantifying,williams2010nonnegative}. Despite this common foundation, the three approaches differ fundamentally in the type of dynamic HOIs they characterize: OIR quantifies undirected HOIs through the balance between redundancy and synergy; PIRD identifies directed redundant and synergistic interactions from multiple sources to an assigned target; PDGC is both directed and directional, quantifying redundant and synergistic Granger-causal effects exerted by multiple source processes on a target.
The conceptual differences and operational similarities between PIRD and PDGC are examined in the Supplementary Material (Sect. I). Both approaches rely on the PID redundancy lattice and decompose quantities that are closely related for Gaussian processes, namely the multivariate mutual information rate (MIR) and the full GC. Their main distinction lies in the symmetric nature of MIR versus the causal nature of GC: PIRD captures bidirectional and instantaneous interdependencies, whereas PDGC extracts causal effects from dynamic symmetric HOIs (Supplementary Material, Sect. IIb).}

\textcolor{black}{Beyond spectral approaches, PDGC is also related to recently introduced methods for analyzing high-order directed interactions, including the Synergistic-Unique-Redundant Decomposition (SURD \cite{martinez2024decomposing}) and the partial causality decomposition of the maximal average causal effect (MACE, \cite{jansma2026decomposing}). Being based on cross-prediction, SURD assesses dynamic directed HOIs but does not reliably identify Granger-causal effects, since cross-prediction methods may infer spurious causality in the reverse direction of a true causal mechanism in the presence of self-dependencies \cite{faes2015information}. Moreover, SURD decomposes directed effects into a reduced number of information atoms (e.g., excluding pairwise redundancies between sources); compared to PID-based methods like ours, this choice improves scalability with the number of sources $N$, but unavoidably conflates contributions from different sources and may thus oversimplify the characterization of the interaction structure.
On the other hand, the MACE decomposition is conceptually closer to PDGC because it also exploits the PID lattice and M\"{o}bius inversion. However, it operates within the fundamentally different framework of interventional causality \cite{pearl2010introduction} applied to random variables. While interventional measures capture genuine physical causation, they are difficult to apply in many real-world settings where only observational data are available. Overall, the recent emergence of diverse methods for decomposing directed and causal interactions illustrates the flexibility of information decomposition, whose formulation ultimately depends on the objectives of the analysis and the characteristics of the available data.}

\textcolor{black}{The application of PDGC to physiological time series revealed the unique, redundant, and synergistic interplay among whole-band and frequency-specific cardiovascular and cerebrovascular oscillations involved in the short-term regulation of heart rate and cerebral blood velocity. It further demonstrated how these information components shape the response to orthostatic stress in healthy subjects and patients prone to syncope.
Our findings reveal markedly different physiological adaptations of the CV and CB systems to orthostatic stress in healthy controls and syncope patients. In healthy subjects, postural stress enhanced CV interactions by increasing the causal influence of SAP on HP variability, whereas this response was markedly attenuated in patients. Conversely, CB interactions changed only in syncope patients, through a tilt-induced increase in the causal effects of HP and MAP on MCBV.
Although these contrasting responses could already be inferred from the multivariate GC computed at rest and during tilt, consistently with previous studies \cite{ocon2009increased,medow2014altered,faes2013mechanisms,pernice2022spectral}, high-order causal analysis revealed them more clearly and linked them to specific information components and frequency bands. In particular, the physiological baroreflex response to postural stress was characterized by unique information transfer from SAP to HP across all frequency bands, together with redundant and synergistic interactions involving respiration in the LF band. In contrast, the pathological cerebrovascular response was characterized by unique, redundant, and synergistic cardiovascular interactions confined to the LF band and independent of respiration.
It should be noted that, even if the significance of the GPDC metrics was assessed by  surrogate-based testing, the relatively small sample size (13 participants per group) limits the statistical power of the group-differences assessed. Nevertheless, although preliminary, our findings suggest that decomposing multivariate interactions among physiological oscillations provides valuable insight into both normal regulatory mechanisms and pathological adaptations.}

\section{Conclusions}

This work combines the frameworks of spectral Granger causality and partial information decomposition, opening the way to the assessment of high-order interdependencies in networks of oscillatory processes. 
The main features of the proposed approach are the flexibility and computational reliability provided by the VAR and SS model implementation of the PDGC measures, together with their intrinsic data-driven nature and the useful possibility of working in the frequency domain to focus the analysis on the oscillatory content of the analyzed processes.
Applied to computational physiology, PDGC revealed high-order patterns of causal interaction among cardiovascular, cerebral, and respiratory time series that may serve as clinical biomarkers of impaired baroreflex control and blood flow dynamics in patients with autonomic dysfunction. More broadly, the proposed method provides a general framework for investigating complex dynamic networks with node activity described by rhythmic processes in computational neuroscience, biology, climatology and other application domains.

%\nocite{*}
\bibliography{biblio.bib}

\end{document}